\documentclass[12pt]{iopart}

\usepackage{units}
\usepackage{epsfig}

\newcommand{\bq}{\begin{equation}}
\newcommand{\eq}{\end{equation}}

\newcommand{\bqq}{\begin{eqnarray}}
\newcommand{\eqq}{\end{eqnarray}}

\newcommand{\cmsNN}{\sqrt{s_{NN}}}

\newcommand{\expval}[1]{\langle #1 \rangle}
\newcommand{\etain}[1]{$|\eta|$~$<$~$#1$}

\renewcommand{\pt}{\ensuremath{p_T} }
\newcommand{\pta}{\ensuremath{p_{T, \rm assoc}}}
\newcommand{\ptt}{\ensuremath{p_{T, \rm trig}}}

\newcommand{\Dphi}{\Delta\phi}
\newcommand{\Deta}{\Delta\eta}
\newcommand{\Ntrig}{N_{\rm trig}}
\newcommand{\Nassoc}{N_{\rm assoc}}
\newcommand{\icp}{I_{CP}}
\newcommand{\iaa}{I_{AA}}

\newcommand{\figref}[1]{Fig.~\ref{#1}}

\newcommand{\bfigFullPage}{\begin{figure} \begin{center} \vspace{0pt}}
\newcommand{\bfig}[1][t!]{\begin{figure}[#1] \begin{center}}
\newcommand{\efig}{\end{center} \end{figure}}
\newcommand{\btab}[1][t!]{\begin{table}[#1] \begin{center}}
\newcommand{\etab}{\end{center} \end{table}}

\newcommand{\eqref}[1]{(\ref{#1})}

\begin{document}

\title{Hadron Correlations in Pb--Pb collisions at $\cmsNN = \unit[2.76]{TeV}$ with ALICE}

\author{Jan Fiete Grosse-Oetringhaus for the ALICE collaboration}

\address{CERN, 1211 Geneva 23, Switzerland}
\ead{Jan.Fiete.Grosse-Oetringhaus@cern.ch}
\begin{abstract}
Untriggered di-hadron correlations studies are shown which provide a map of the bulk correlation structures in Pb-Pb collisions. Long-range correlations are further studied by triggered correlations which address the dependence on trigger and associated $\pt$. Measured correlation functions are decomposed with a
multi-parameter fit and into Fourier coefficients. The jet-yield
modification factor $\iaa$ is
presented.
\end{abstract}

The objective of the study of ultra-relativistic heavy ion-collisions  is the characterization of
the quark--gluon plasma, the deconfined state of quarks and gluons. Recent measurements by ALICE
indicate that in central Pb-Pb collisions at the LHC unprecedented color charge densities are
reached. For example, the suppression of charged hadrons in central collisions
expressed as the nuclear modification factor $R_{\rm AA}$ as a function of
transverse momentum ($\pt$) reaches a value as low as about 0.14 \cite{raa}.

Di-hadron correlations allow for the further study of in-medium energy loss because for most pairs of
partons scattered in opposite directions, one will have a longer path through the medium than the
other. Two-particle correlations allow the study of medium effects on the jet fragmentation without the need of jet
reconstruction. This paper presents results from untriggered correlations where all charged
particles are correlated with each other providing a map of the bulk correlation structures. A long-range correlation structure on the near-side, usually called the `ridge'
\cite{starridge}, is discussed and its compatibility with collective flow studied.
Triggered correlations are studied to assess at which $\pt$ collective effects dominate
and in which region jet-like correlations contribute most. In the jet-dominated
regime, at higher $\pt$, near- and away-side yields are measured and compared between
central and peripheral events ($\icp$) and studied relative to a pp reference ($\iaa$).

\section{Detector and Data Sample}

The Inner
Tracking System and the Time Projection Chamber (TPC) of the ALICE detector \cite{alice} are used for vertex finding and
tracking. Forward scintillators (VZERO) determine the centrality of the collisions.
The main tracking information stems from the TPC which has a uniform acceptance in azimuthal angle ($\phi$) and a pseudorapidity coverage of
\etain{0.9}. The reconstructed vertex information is used to select primary track candidates and to constrain the $\pt$ of the track.
The uniform acceptance results in only small required corrections: the $\phi$-distribution of tracks is rather flat  and the mixed-event correction as function of the difference in $\eta$ and $\phi$ of two particles ($\Deta$ and $\Dphi$) is close to a perfect triangle.
About 14 million minimum-bias events recorded in fall 2010 have been used in the analysis.
Good-quality tracks are selected by requiring at least 70 (out of maximally 159) associated clusters in the
TPC, and a $\chi^2$ per space point of the momentum fit smaller than 4. In addition, tracks are
required to originate from within \unit[2.4]{cm} (transverse) and \unit[3.2]{cm} (longitudinal) of the primary vertex.

\section{Untriggered Correlations}

\bfig
  \includegraphics[width=\linewidth,trim=0 15 0 0,clip=true]{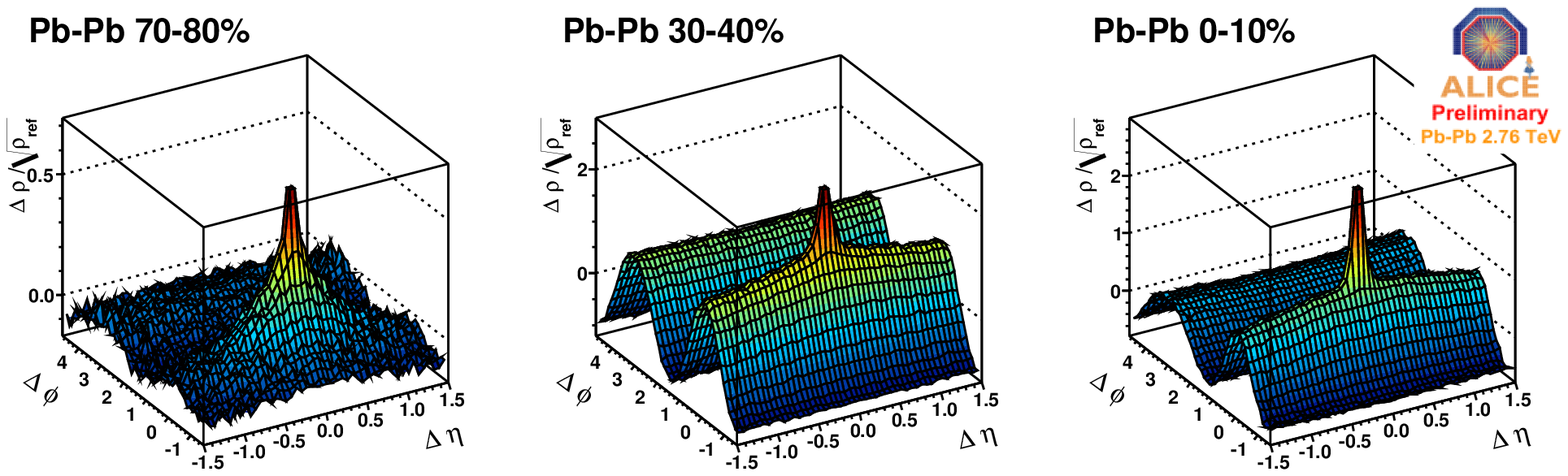}
  \caption{\label{untriggered_centrality} Centrality evolution from peripheral (left) to central (right) collisions of the untriggered correlation function.}
\efig

In untriggered correlations all charged particles ($\pt > \unit[0.15]{GeV/\emph{c}}$) are correlated with each other. The correlation is expressed as:
\bq
    \frac{\Delta \rho(\Deta, \Dphi)}{\sqrt{\rho_{\rm ref}(\Deta, \Dphi)}} =
      \frac{\rho_{\rm same} - \rho_{\rm ref}}{\sqrt{\rho_{\rm ref}}} =
      \frac{dN_{\rm ch}}{d\eta d\phi} \left( \frac{\rho_{\rm same}}{\rho_{\rm ref}} - 1 \right) \label{eq_untriggered}
\eq
where $\rho_{\rm same}$ and $\rho_{\rm ref}$ is the two-particle correlation as function of $\Deta$ and $\Dphi$ of particles from same and different events, respectively, and $dN_{\rm ch}/d\eta$ the average multiplicity.
Eq.~\eqref{eq_untriggered} explicitly contains the average number of charged particles and therefore is a measure of the number of correlated pairs per particle.
\figref{untriggered_centrality} shows its centrality evolution.
At all centralities a sharp peak around $\Deta \approx 0$ and $\Dphi \approx 0$ can be seen attributed to conversions and Bose--Einstein correlations of identical particles.
The correlation structures increase with centrality (note the different scales). Mid-central collisions are dominated by elliptic flow ($\cos 2 \Delta \phi$ structure). Further, a $\Deta$-dependent long-range correlation structure at $\Dphi \approx 0$ builds up towards central collisions.

\bfig
  \includegraphics[width=0.95\linewidth,trim=0 5 0 5,clip=true]{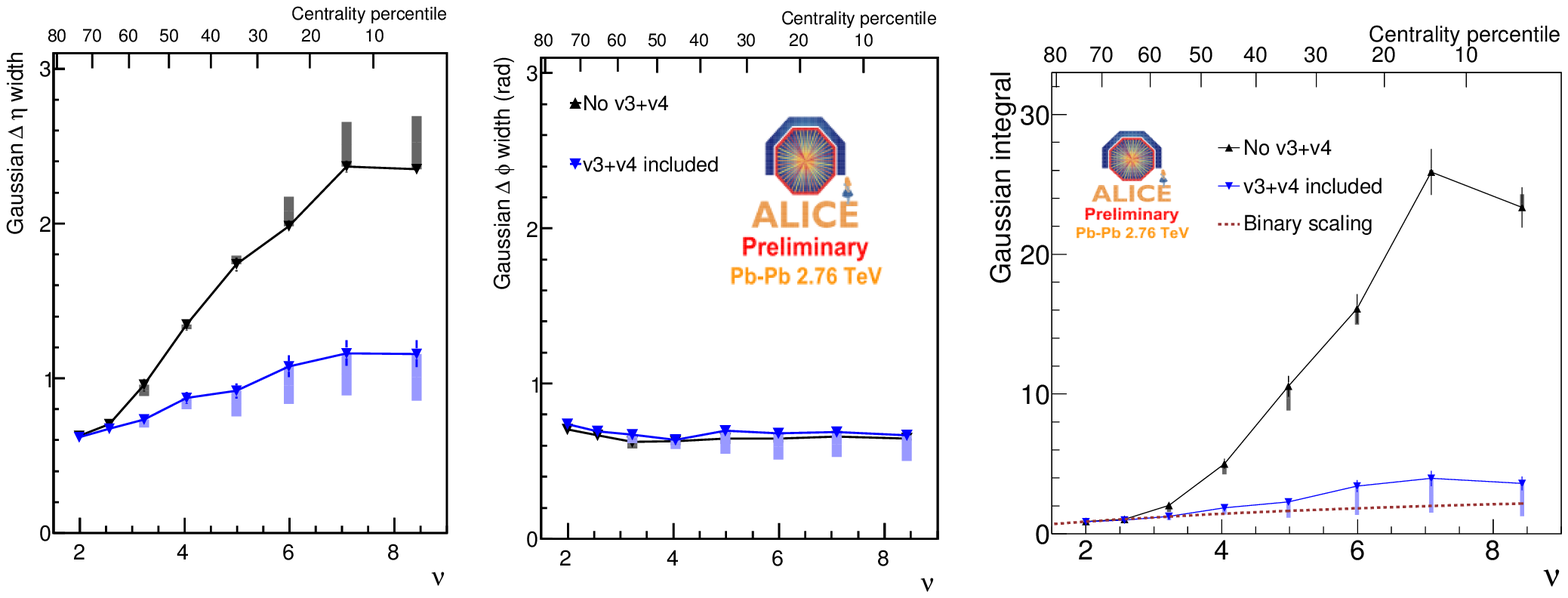}
  \caption{\label{untriggered_parameters} The fit parameters $\sigma_{\Deta}$ (left), $\sigma_{\Dphi}$ (center), and Gaussian volume (right) as function of centrality and $\nu = 2 \expval{N_{\rm coll}} / \expval{N_{\rm part}}$.}
\efig

The correlation function is fitted with a combination of a two dimensional Gaussian with different
widths in $\Deta$ and $\Dphi$ centered at $\Dphi = 0$ and $\Deta = 0$ and four $\cos n \Dphi$ terms describing $\Deta$
independent effects:
\bq
    \frac{\Delta \rho}{\sqrt{\rho_{\rm ref}}} = A + B Gauss_{\sigma_{\Deta},\sigma_{\Dphi}}(\Deta, \Dphi) + \sum_{n=1}^4 C_n \cos n \Dphi.
\eq
The fit is performed with all four $C_n$ coefficients as well as with only the first two (similar to earlier measurements \cite{staruntrig}). Both ways describe the correlation well with a $\chi^2/{\rm ndf} \approx 1.3 - 1.7$. The long-range $\Deta$-independent correlation structure around $\Dphi \approx 0$ seems therefore to be of collective origin.
In addition to the first four $C_n$ coefficients which cause by definition no $\Deta$-modulation, a near-side structure is present with Gaussian widths of $\sigma_{\Deta} \approx 0.6 - 1.0$ and $\sigma_{\Dphi} \approx 0.6 - 0.7$.
\figref{untriggered_parameters} shows $\sigma_{\Deta}$ and $\sigma_{\Dphi}$ as function of centrality: $\sigma_{\Deta}$ is significantly larger when $C_3$ and $C_4$ are not included; in both cases it increases towards central collisions. $\sigma_{\Dphi}$ is rather independent of centrality.
The right panel of \figref{untriggered_parameters} shows the Gaussian volume $2 \pi B \sigma_{\Deta} \sigma_{\Dphi}$, i.e. the total number of correlated pairs per particle contained in the structure. It increases with centrality and
is compatible with binary scaling which assumes that the number of correlated pairs scales with the number of binary collisions (when $C_3$ and $C_4$ are included). Further this structure shows a strong charge dependence: \figref{untriggered_charge} shows the correlation function for unlike-sign and like-sign pairs. The like-sign structure is much wider, while the unlike-sign structure contributes most pairs to the Gaussian which could mostly rise from jet-like correlations of a few particles.
More details can be found in \cite{anthonyproceedings}.

\bfig
  \includegraphics[width=0.7\linewidth,trim=0 20 0 0,clip=true]{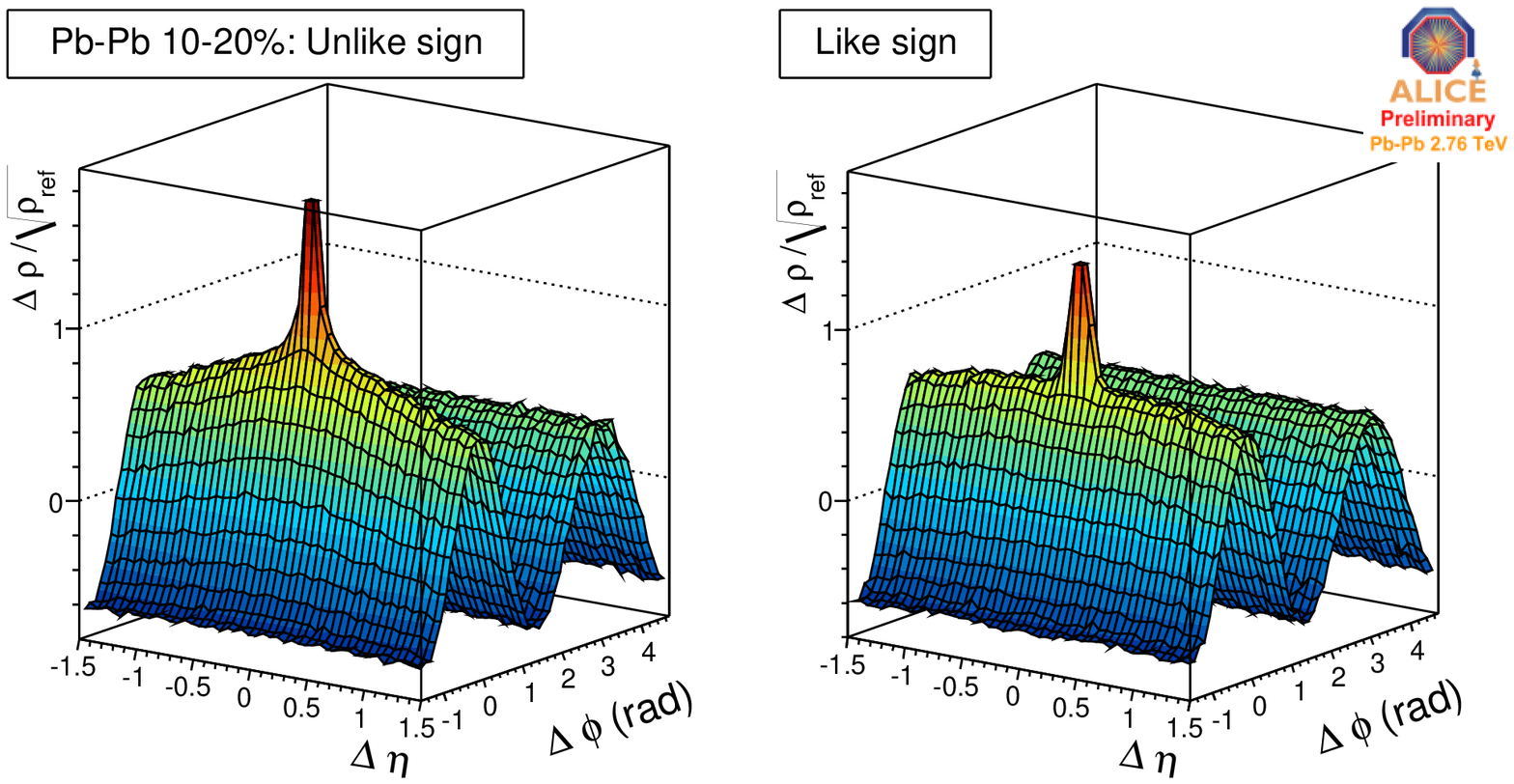}
  \caption{\label{untriggered_charge} The unlike-sign (left) and like-sign (right) correlation function.}
\efig

\section{Fourier Decomposition}

The $\pt$ dependence of the correlation is studied by measuring triggered correlations. In such an analysis a particle is chosen from a $\pt$ region and called \emph{trigger particle}. So-called \emph{associated particles} from another $\pt$ region are correlated to the trigger particle where $\pta < \ptt$. The correlation function is defined as:
\bq
  C(\Deta, \Dphi) = \left( \frac{1}{N_{\rm pairs}} \frac{d\Nassoc}{d\Dphi d\Deta} \right)_{\rm same} / \left( \frac{1}{N_{\rm pairs}} \frac{d\Nassoc}{d\Dphi d\Deta} \right)_{\rm mixed}
\eq
where $\Ntrig$ is the number of trigger particles to which $\Nassoc$ particles are associated at $\Dphi = \phi_{\rm trig} - \phi_{\rm assoc}$ and $\Deta = \eta_{\rm trig} - \eta_{\rm assoc}$, and $N_{\rm pairs}$ the total number of pairs. The subscript same (mixed) indicates that pairs are formed from the same (different) events.

\bfig
  \hspace{0.03\linewidth}
  \includegraphics[width=0.43\linewidth,trim=0 5 0 10,clip=true]{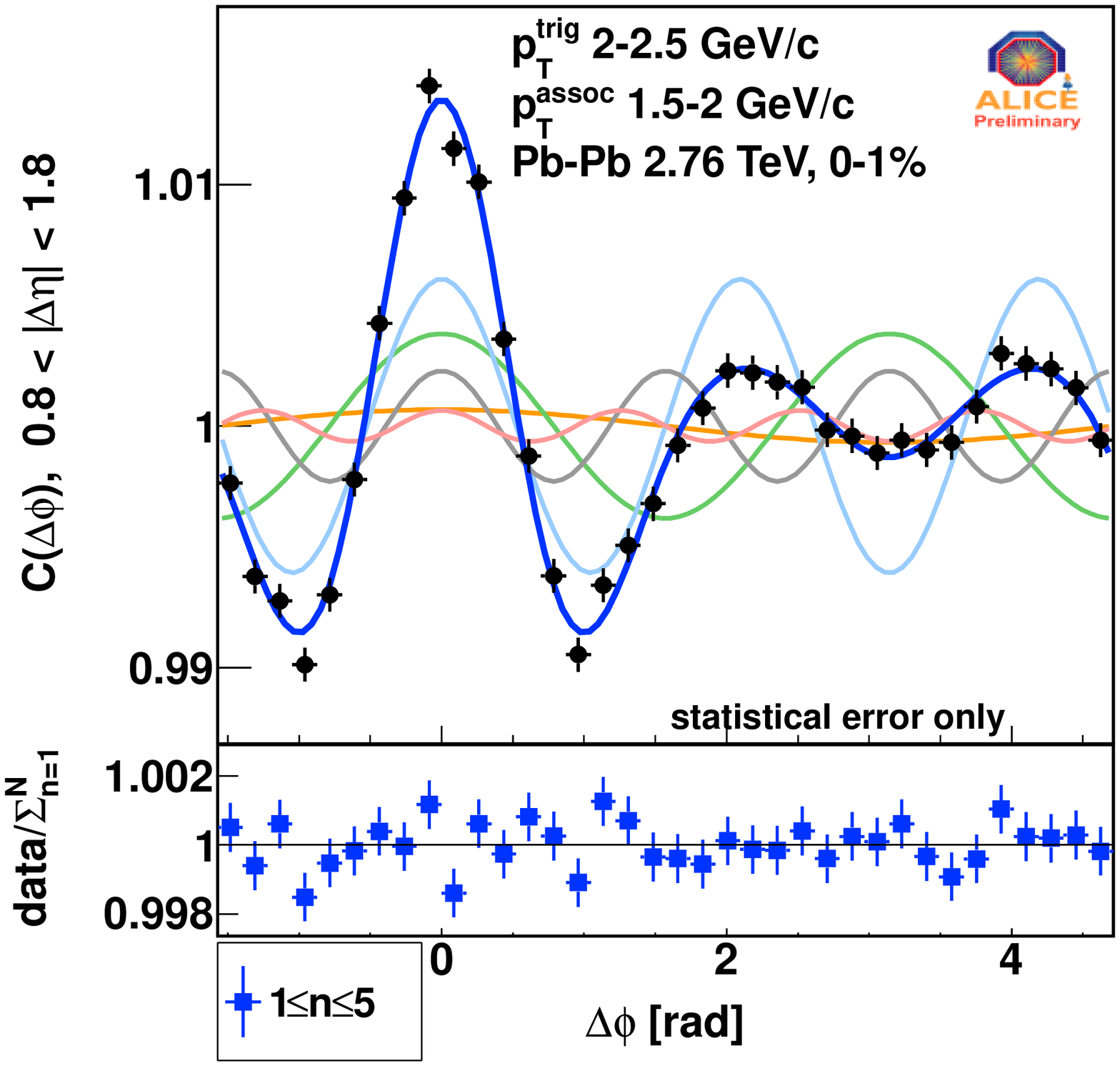}
  \hfill
  \includegraphics[width=0.43\linewidth,trim=0 5 0 10,clip=true]{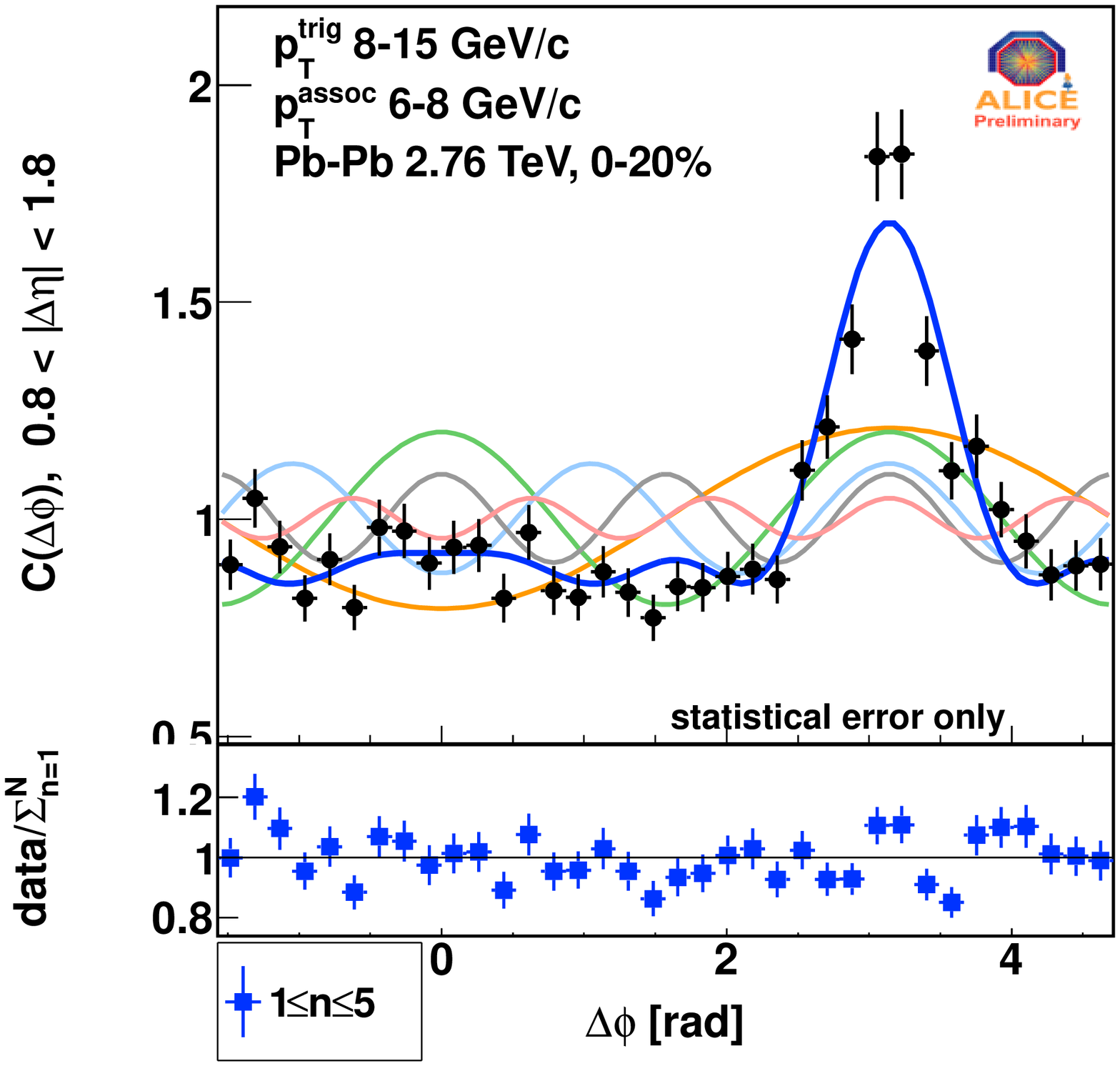}
  \hspace{0.03\linewidth}
  \caption{\label{fourier_examples} Correlation function (points) and its Fourier decomposition (lines) in $0.8 < |\Deta| < 1.8$ for a low $\pt$ bin in \unit[0-1]{\%} centrality (left) and a high $\pt$ bin (left) in \unit[0-20]{\%} centrality (right). The $\pt$ ranges are indicated on the plot.
  The first five Fourier harmonics are shown as well as their sum. The lower part shows the residuals between the  data and the sum of the harmonics.}
\efig

\bfig
  \includegraphics[width=0.9\linewidth,trim=0 15 0 5,clip=true]{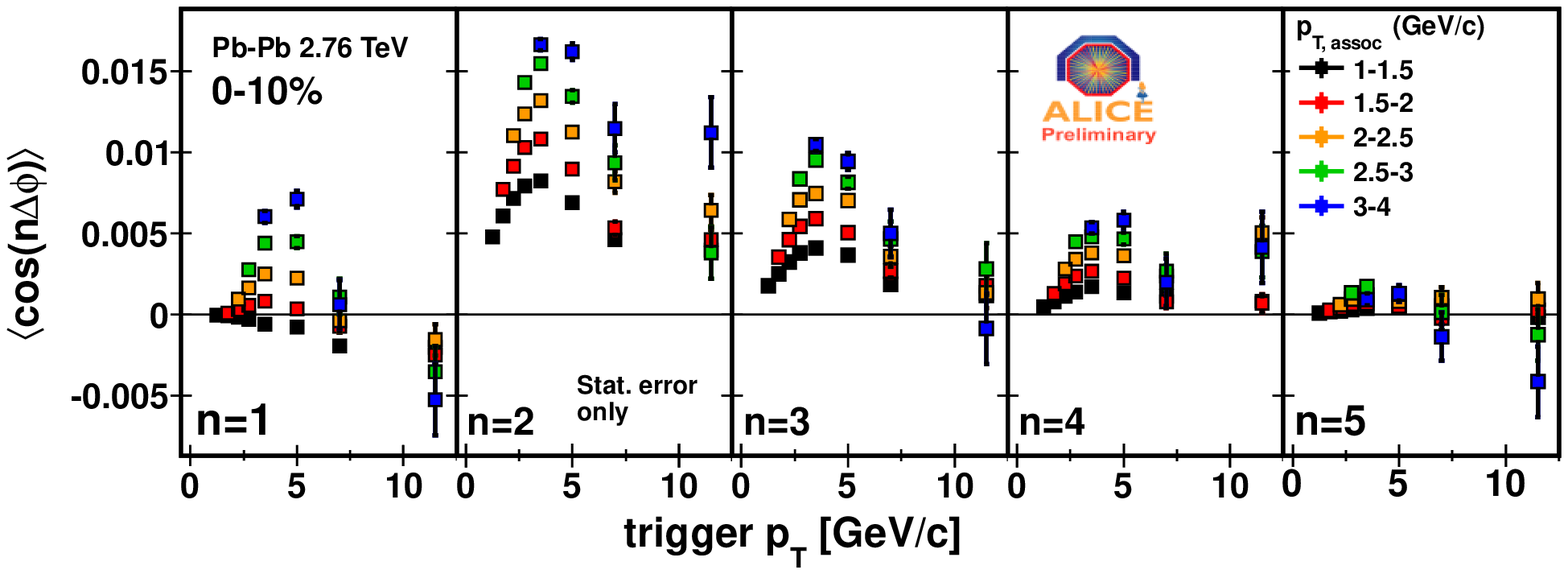}
  \caption{\label{fourier_coefficients} Fourier coefficients for \unit[0-10]{\%} centrality. The x-axis denotes $\ptt$ and the different symbols different ranges of $\pta$.}
\efig

The long-range correlation region within $0.8 < |\Deta| < 1.8$ is projected to $\Dphi$ and decomposed into Fourier coefficients. These are calculated with:
\bq
  V_{n\Delta} = \expval{\cos n \Dphi} = \frac{\int d\Dphi C(\Dphi) \cos n \Dphi}{\int d\Dphi C(\Dphi)}.
\eq
\figref{fourier_examples} shows $C(\Dphi)$ in a low $\pt$ and a high $\pt$ bin. The first 5 Fourier coefficients are shown as well as their sum which describes the correlation well at low $\pt$ and reasonably at high $\pt$. At low $\pt$, no significant improvement is observed by including higher coefficients. In the low $\pt$ bin a double-humped structure on the away-side is observed (notably without any background subtraction) which is discussed in more detail in \cite{newflow, andrewproceedings}. The extracted coefficients for \unit[0-10]{\%} centrality are shown in \figref{fourier_coefficients}. They increase with increasing $\pt$ and drop for $\ptt > \unit[5]{GeV/\emph{c}}$. The odd terms become negative at large $\pt$ which is attributed to the influence of the away-side jet. From central to peripheral collisions, $V_{2\Delta}$ rises most (plots not shown).
To assess which part of the correlation is caused by collective effects, we test the factorization relation:
$V_{n\Delta}(\pta,\ptt) \stackrel{?}{=} v_{n}(\pta) \cdot v_{n}(\ptt)$
which is valid if the two-particle correlation is connected by a common plane of symmetry. This is not the case for jet-like correlations where a few particles are correlated by fragmentation\footnote{Indirect correlations exist, e.g. length-dependent quenching which has the largest influence on $V_{2\Delta}$.}.
In addition to the coefficients presented above which are extracted as function of $\pta$ and $\ptt$, i.e. $V_{n\Delta}(\pta,\ptt)$, a global fit is performed allowing only one value per $\pt$-bin, i.e. $v_{n}(\pt)$.
\figref{fourier_factorization} presents the test of the factorization relation for $V_{2\Delta}$ and \unit[0-10]{\%} centrality. The bottom part of the figure shows the ratio between the two indicating that the factorization works well at low to intermediate $\pt$ (up to \unit[3-4]{GeV/\emph{c}} depending on centrality) and breaks down above where jet-like correlations dominate.
This trend is followed for $n>2$ as well, but $V_{1\Delta}$ does not seem to follow a clear factorization pattern even in flow-dominated regimes (plots not shown).
In the $\pt$ region where the factorization holds, the extracted coefficients are consistent with other measurement prescriptions, e.g. \cite{newflow}.
More details can be found in \cite{andrewproceedings}.

\bfig
  \begin{minipage}[t]{0.58\linewidth}
    \vspace{0pt}
    \centering
    \includegraphics[width=0.9\linewidth,trim=0 20 0 5,clip=true]{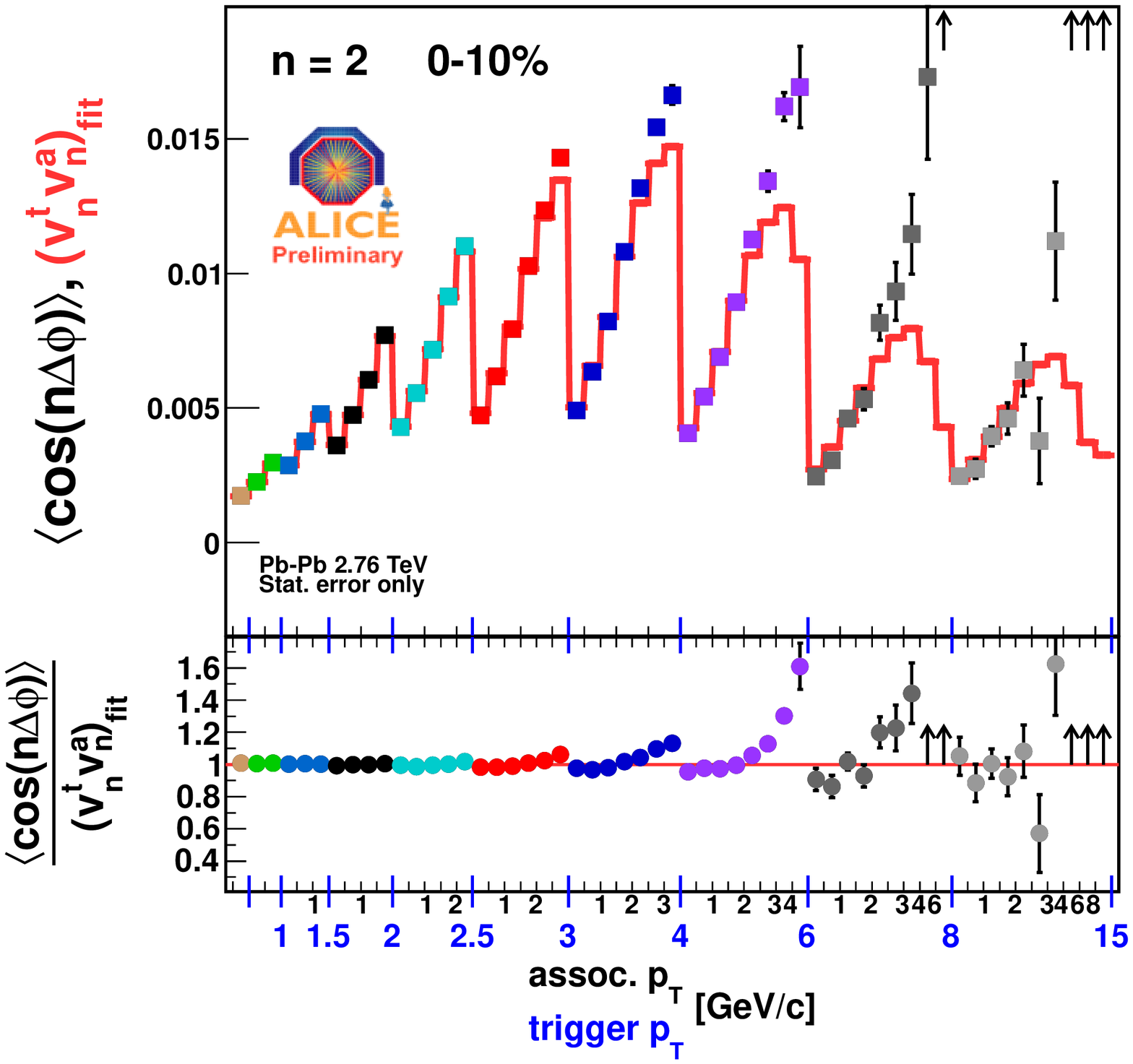}
    \caption{\label{fourier_factorization} Test of the factorization relation for: $V_{2\Delta}(\pta,\ptt)$ (points); $v_{2}(\pta) \cdot v_{2}(\ptt)$ (line). The bottom panel shows their ratio.}
  \end{minipage}
  \hfill
  \begin{minipage}[t]{0.40\linewidth}
    \vspace{0pt}
    \centering
    \includegraphics[width=0.9\linewidth,trim=0 5 0 5,clip=true]{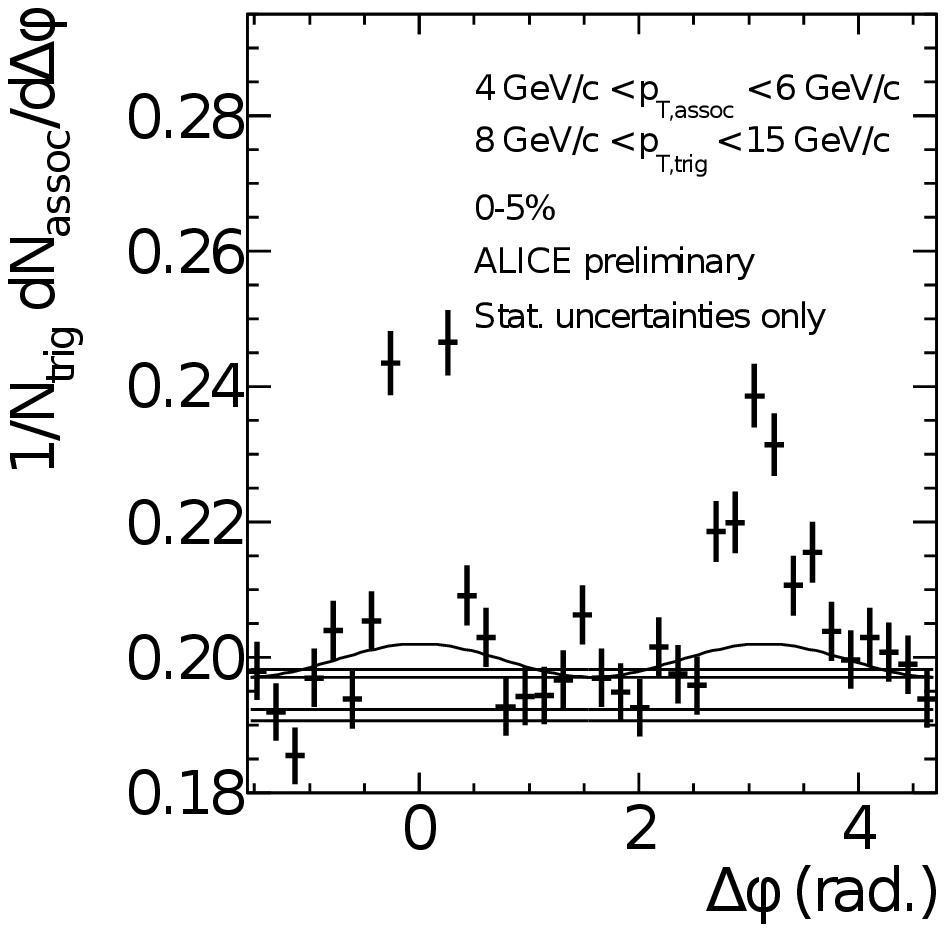}
    \caption{\label{fig_pedestal} Per-trigger yield (normalized by $\Deta = 1.6$) in an example bin (zoom, the near-side peak is off scale). The determined pedestal values (horizontal lines) and the $v_2$ component ($\cos 2 \Dphi$) are shown.}
  \end{minipage}
\efig

\section{Modification of the Jet-Particle Yield}

At higher $\pt$ where collective effects are small and jet-like correlations dominate ($\ptt > \unit[8]{GeV/\emph{c}}$; $\pta > \unit[3]{GeV/\emph{c}}$), the effect of the medium on the yield of particles in a jet has been studied. This is assessed by calculating ratios of yields on the near- and away-side. To remove uncorrelated background from the associated yield, the pedestal value needs to be determined. This is done by fitting the region close to the minimum of the $\Dphi$ distribution ($\Dphi \approx \pm \frac{\pi}{2}$) with a constant and using this value as pedestal (zero yield at minimum -- ZYAM). One cannot exclude a correlated contribution in this region (e.g. from 3-jet events), and we do not claim that we only remove uncorrelated background. Instead we measure a yield with the prescription given here.
To estimate the uncertainty on the pedestal determination, we use four different approaches (different fit regions as well as averaging over a number of bins with the smallest content). \figref{fig_pedestal} shows the per-trigger yield for an example bin. The horizontal lines indicate the determined pedestal values; their spread gives an idea of the uncertainty. Also indicated is a background shape considering $v_2$. The $v_2$ values are taken from an independent measurement \cite{newflow}; for the centrality class \unit[60-90]{\%} no $v_2$ measurement was available, therefore, as an upper limit, $v_2$ is taken from the \unit[40-50]{\%} centrality class as it is expected to reduce towards peripheral collisions). For a given bin the $v_2$ background is $2\expval{v_{\rm 2,trig}}\expval{v_{\rm 2,assoc}} \cos 2 \Dphi$ where the $\expval{...}$ is calculated taking into account the $\pt$ distribution of the trigger and associated particles.
Subsequently to the pedestal (and optionally $v_2$) subtraction, the near- and away-side yields are integrated within $\Dphi$ of $\pm 0.7$ and $\pi \pm 0.7$, respectively.

\bfig
  \includegraphics[width=0.9\linewidth,trim=0 36 0 12,clip=true]{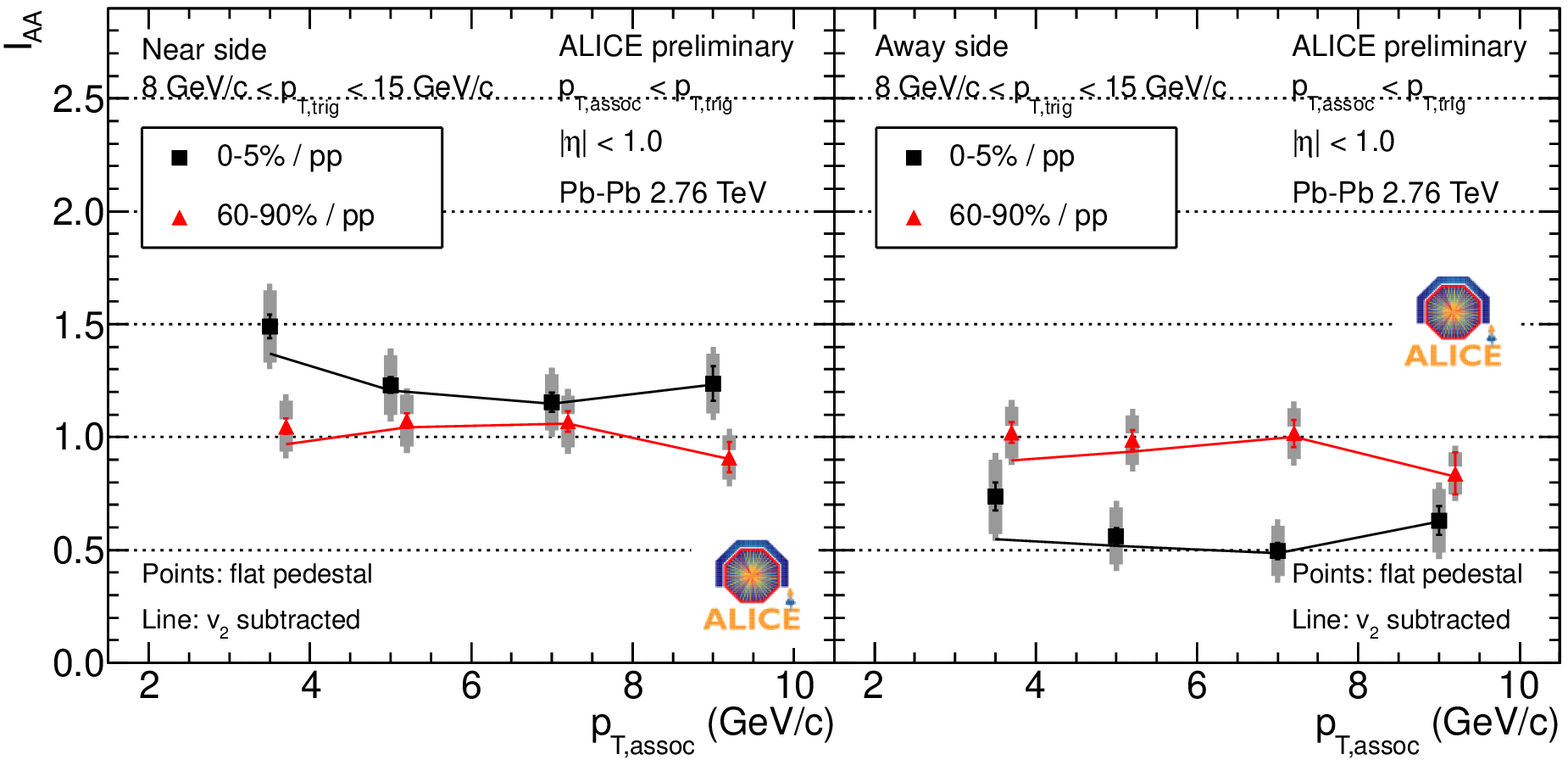}
  \includegraphics[width=0.9\linewidth,trim=0 15 0 14,clip=true]{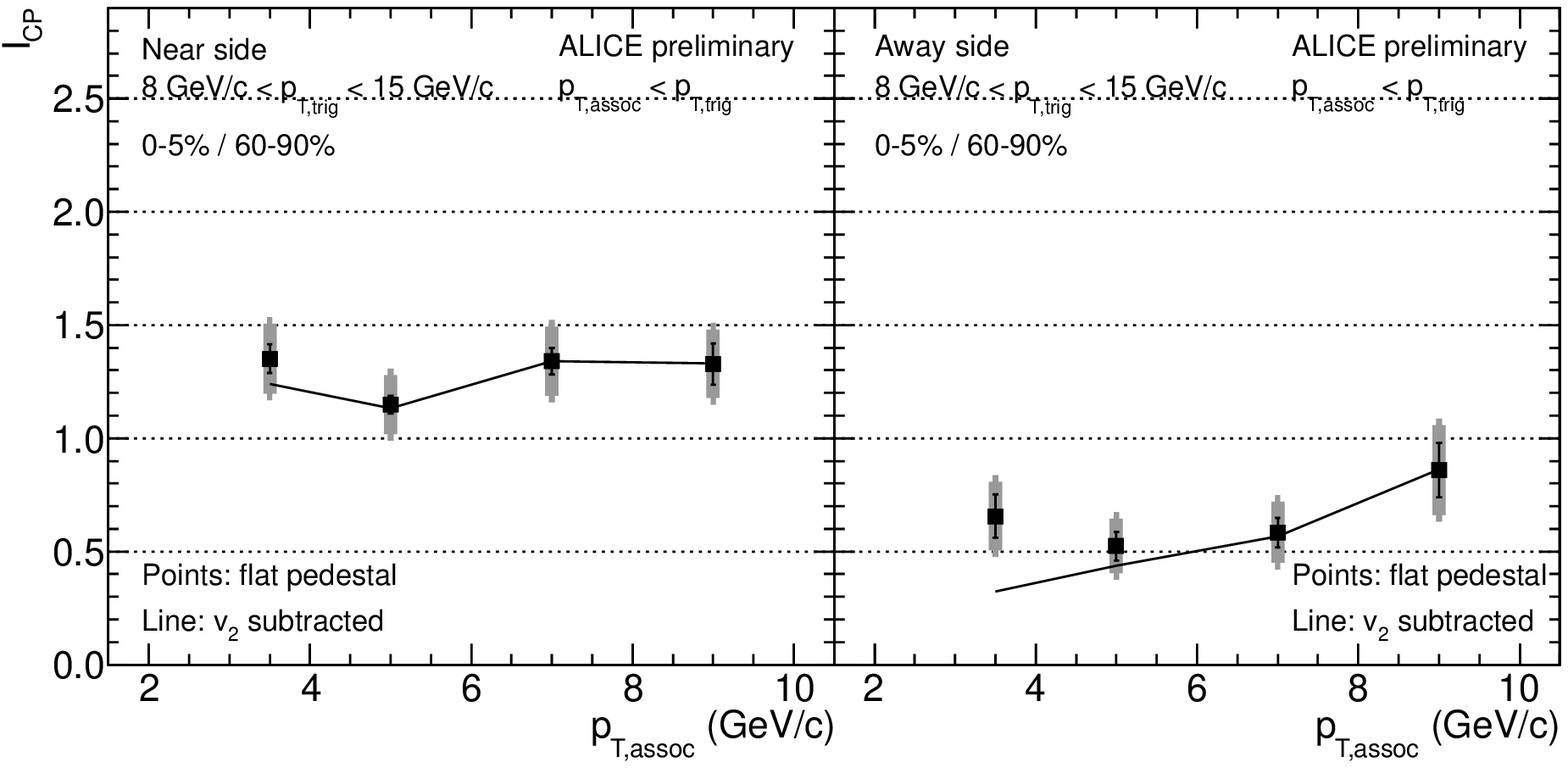}
  \caption{\label{fig_iaaicp} $\iaa$ (top panel) and $\icp$ (bottom panel): the data points are calculated with a flat pedestal; the line is based on $v_2$ subtracted yields.}
\efig

To quantify the effect of the in-medium energy loss, the ratio of yields in Pb-Pb and pp collisions is calculated:
\figref{fig_iaaicp} (top panel) shows $\iaa$ for central and peripheral collisions using the flat pedestal (data points) and $v_2$ subtracted yields (lines).
The only significant difference is in the lowest bin of $\pta$ which confirms the small bias due
to flow anisotropies in this $\pt$ region.
Note that only $v_2$ is considered here, although contributions from other harmonics might be of the same order, particularly for central events. In central collisions, away-side suppression from in-medium energy loss is seen ($\iaa \approx 0.6$), as expected. Moreover, there is an unexpected enhancement above unity ($\iaa \approx 1.2$) on the near-side that was not observed with significance at lower collision energies \cite{iaastar}. In peripheral collisions, both near- and away-side are consistent with unity.

Further, the ratio of yields in central and peripheral collision, $\icp$, has been calculated,
shown in \figref{fig_iaaicp} (bottom panel).
The difference is rather small and only in the smallest $\pta$ bins. $\icp$ is consistent with $\iaa$ in central collisions with respect to the near-side enhancement and the away-side suppression.

It is for the first time that a significant near-side enhancement of $\iaa$ and $\icp$ is observed showing that the near-side parton is also quenched.
For LHC energies an enhancement of \unit[10-20]{\%} was predicted albeit for larger $\ptt$
\cite{Renk:2007rn}. It is attributed to the enhanced relative abundance of quarks w.r.t. gluons
escaping the medium: gluons couple stronger to the medium due to their different color charge and
their abundance is reduced. The quarks fragment harder and thus produce an enhanced associated
yield. In addition, energy loss of the near-side parton causes that trigger particles with similar $\pt$ stem from partons with higher
$\pt$ in Pb-Pb collisions than in pp collisions. Consequently, more energy is available for
particle production on near- and away-side.

\bfig
  \includegraphics[width=0.87\linewidth,trim=0 15 0 12,clip=true]{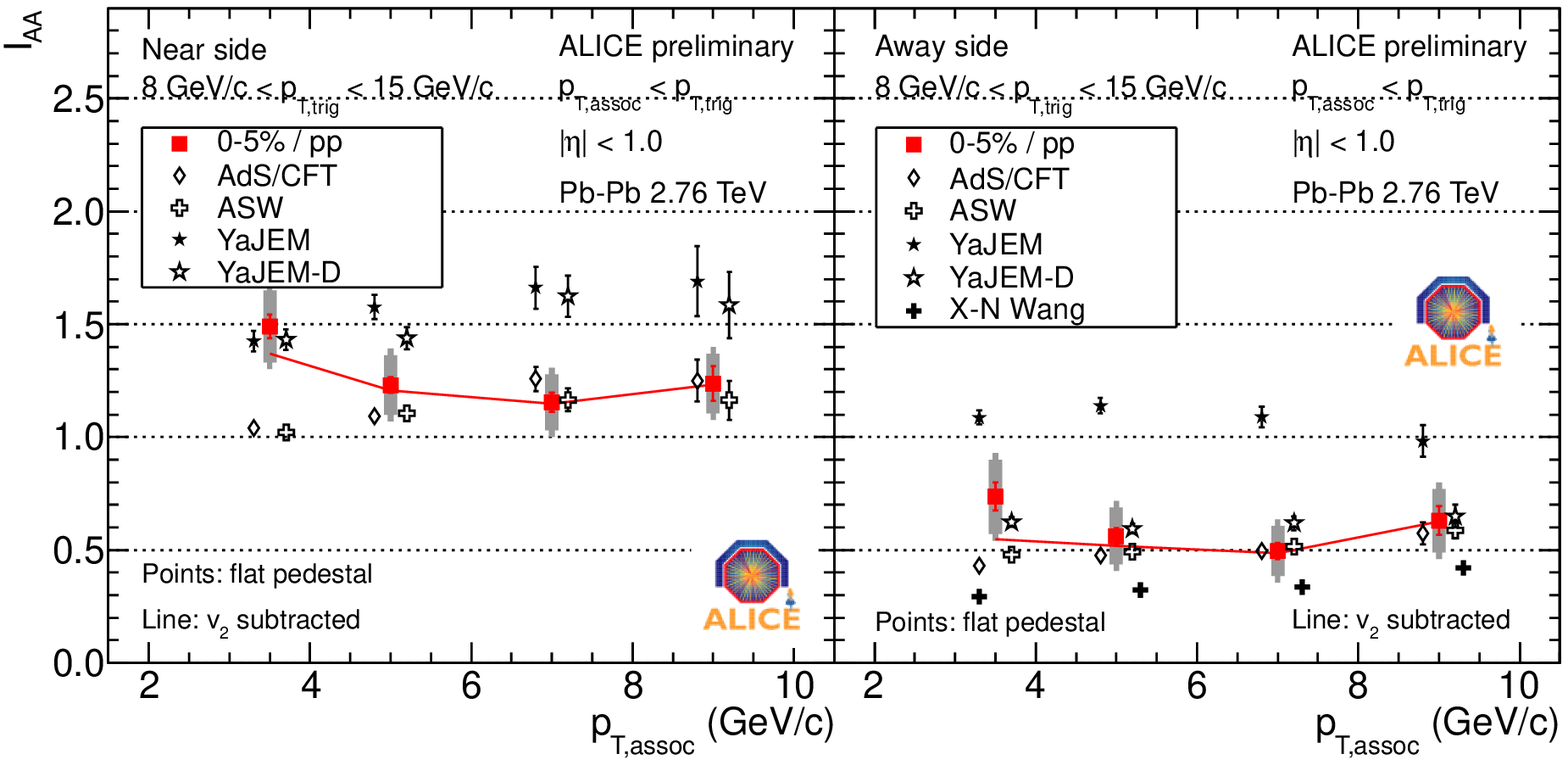}
  \caption{\label{fig_iaatheory} $\iaa$ compared to theoretical calculations.}
\efig

\figref{fig_iaatheory} shows $\iaa$ compared to theory predictions. Shown are calculations with different energy loss scenarios \cite{renkprediction,wangprediction}. The near-side enhancement is reproduced by energy loss following AdS/CFT (cubic path length dependence) and ASW (quadratic dependence). YaJEM (linear dependence) as well as YaJEM-D yield too large values. The away-side suppression is reproduced by AdS/CFT, ASW and YaJEM-D; YaJEM is too high and the calculation from X.~N.~Wang yields slightly too low values.

\section{Summary}

An extensive study of dihadron correlations in Pb-Pb collisions at $\cmsNN = \unit[2.76]{TeV}$ measured with ALICE has been presented. We show that the near-side long-range correlation structure (known as the `ridge') is compatible with collective flow and `only' a two-dimensional Gaussian structure with a width of about $0.6-1$ remains after flow has been considered.
We decompose the long-range correlation region into Fourier coefficients and study their factorization relation showing that at low to intermediate $\pt$ (up to \unit[3-4]{GeV/\emph{c}}) the correlation function is dominated by collectivity while at larger $\pt$ jet-like correlations dominate. In the latter region we measure modification of the jet-particle yield and find away-side suppression consistent with strong in-medium energy loss as well as an interesting near-side enhancement showing that the effect of the medium on the near-side is measurable at the LHC.

\end{document}